\documentclass[pdflatex,sn-mathphys-num]{sn-jnl}% Math and Physical Sciences Numbered Reference Style 
\usepackage{graphicx}%
\usepackage{multirow}%
\usepackage{amsmath,amssymb,amsfonts}%
\usepackage{amsthm}%
\usepackage{mathrsfs}%
\usepackage[title]{appendix}%
\usepackage{xcolor}%
\usepackage{textcomp}%
\usepackage{manyfoot}%
\usepackage{booktabs}%
\usepackage{algorithm}%
\usepackage{algorithmicx}%
\usepackage{algpseudocode}%
\usepackage{listings}%
\usepackage{tcolorbox}
\usepackage{booktabs}
\usepackage{tabularx}
\usepackage{longtable} 

\theoremstyle{thmstyleone}%
\theoremstyle{thmstyletwo}%

\theoremstyle{thmstylethree}%

\begin{document}

\title[Article Title]{Integrating Generative AI in Cybersecurity Education: Case Study Insights on Pedagogical Strategies, Critical Thinking, and Responsible AI Use}

\author*[1]{\fnm{Mahmoud} \sur{Elkhodr}}\email{m.elkhodr@cqu.edu.au}
\author[1]{\fnm{Ergun} \sur{Gide}}\email{e.gide1@cqu.edu.au}

\affil[1]{\orgname{Central Queensland University}, \orgaddress{\country{Australia}}}

%%========================================================
%%


\abstract{The rapid advancement of Generative Artificial Intelligence (GenAI) has introduced new opportunities for transforming higher education, particularly in fields that require analytical reasoning and regulatory compliance, such as cybersecurity management. This study presents a structured framework for integrating GenAI tools into cybersecurity education, demonstrating their role in fostering critical thinking, real-world problem-solving, and regulatory awareness. The implementation strategy followed a two-stage approach, embedding GenAI within tutorial exercises and assessment tasks. Tutorials enabled students to generate, critique, and refine AI-assisted cybersecurity policies, while assessments required them to apply AI-generated outputs to real-world scenarios, ensuring alignment with industry standards and regulatory requirements. Findings indicate that AI-assisted learning significantly enhanced students’ ability to evaluate security policies, refine risk assessments, and bridge theoretical knowledge with practical application. Student reflections and instructor observations revealed improvements in analytical engagement, yet challenges emerged regarding AI over-reliance, variability in AI literacy, and the contextual limitations of AI-generated content. Through structured intervention and research-driven refinement, students were able to recognise AI’s strengths as a generative tool while acknowledging its need for human oversight. This study further highlights the broader implications of AI adoption in cybersecurity education, emphasising the necessity of balancing automation with expert judgment to cultivate industry ready professionals. Future research should explore the long-term impact of AI-driven learning on cybersecurity competency, as well as the potential for adaptive AI-assisted assessments to further personalise and enhance educational outcomes}




\maketitle

\section{Introduction}

The rapid advancement of Generative Artificial Intelligence (GenAI) technologies has reshaped various sectors, including education, by offering new opportunities for instructional innovation. Among these technologies, ChatGPT has emerged as a widely adopted tool with the potential to enhance teaching and learning through personalised assistance, real-time feedback, and enriched student engagement \cite{elkhodr2023ict}. While GenAI has demonstrated substantial promise, its integration into academic settings remains a subject of ongoing debate, particularly concerning ethical considerations, academic integrity, and the potential risks of over-reliance on AI-generated content \cite{almarzouqi2024ethical}. This study presents a structured framework for embedding GenAI tools into cybersecurity management education, illustrating how such technologies can be effectively leveraged to develop critical thinking, contextual understanding, and problem-solving skills.

ChatGPT has already been successfully implemented across various academic disciplines, serving functions such as generating multiple-choice questions \cite{rivera2024exploring}, assisting with programming tasks in languages like Java and Python, and providing language learning feedback \cite{bucaioni2024programming}. Studies suggest that its use in blended learning environments enhances student engagement, fosters creativity, and facilitates reflective thinking through interactive and iterative learning processes \cite{jayasinghe2024promoting, lee2024empowering, wu2024promoting}. These capabilities position GenAI as a transformative educational tool, particularly in cybersecurity management, which demands analytical rigour and adaptability to evolving threats. However, despite its advantages, GenAI integration in education presents notable challenges, including concerns over potential misuse, the accuracy of AI-generated outputs, and the risk of diminishing students' independent problem-solving capabilities \cite{sandu2024role}. These concerns underscore the need for a structured pedagogical framework that ensures responsible and effective deployment of GenAI while maintaining academic integrity.

Cybersecurity management education represents a particularly compelling case for the integration of GenAI due to its dual emphasis on theoretical knowledge and practical application. The discipline requires students to analyse regulatory frameworks, develop security policies, and address emerging cyber threats—tasks that benefit from AI-driven analytical support while also requiring human oversight to ensure accuracy and relevance. This study introduces a two-stage strategy for incorporating GenAI into cybersecurity management education, designed to maximise the benefits of AI tools while mitigating their limitations through structured guidance and critical engagement.

The first stage of this strategy involves embedding GenAI into tutorial exercises, where students interact with AI-generated content, critique its quality, and refine it using academic research and external sources. This process fosters a deeper understanding of cybersecurity concepts by encouraging students to engage in analytical evaluation rather than passive consumption of AI-generated responses. The second stage integrates GenAI into written assessments, requiring students to apply AI-generated outputs to practical cybersecurity scenarios while iteratively refining their work to align with industry standards and regulatory frameworks. To ensure meaningful learning outcomes, students are also required to engage in reflective exercises, critically assessing their use of GenAI and articulating their insights regarding its efficacy and limitations. By documenting and analysing the outcomes of this pedagogical approach, this study provides a replicable framework for integrating GenAI into higher education curricula, offering a roadmap for educators seeking to adopt AI-enhanced teaching methodologies. 

The findings demonstrate how structured AI integration can enhance learning while preserving the essential role of human expertise. This study ultimately advocates for a balanced approach that combines the efficiency of AI-generated insights with the critical judgment and contextual understanding fostered by traditional pedagogical methods, ensuring that students develop both technical proficiency and independent analytical skills in cybersecurity management.

The remainder of this paper is organised as follows. Section 2 reviews related work in AI-assisted education and cybersecurity teaching methodologies. Section 3 details our implementation strategy and instructional design, including the pedagogical framework and specific case studies. Section 4 presents the results of our implementation, analysing student feedback and learning outcomes. Section 5 discusses the effectiveness of our approach, comparing it with traditional methods and providing recommendations for future implementations. Finally, Section 6 addresses the limitations of our study and while Section 7 concludes the paper and suggests directions for future research

%%%%%%%%%%%%%%%%%%%%%%%%%%%%%%%%%%%%%%%%%%
\section{Related Works}
The integration of AI tools in education has garnered significant attention, with several studies exploring their impact on student learning outcomes and pedagogical effectiveness. A comprehensive review of ChatGPT applications in higher education highlighted its potential for enhancing student engagement and promoting active learning across various disciplines \cite{rasul2023role}. Similarly, research examining the effectiveness of GenAI tools in computer science education demonstrated improved learning outcomes through AI-assisted programming exercises and code review processes \cite{cubillos2025generative}.

In cybersecurity education specifically, research has increasingly focused on innovative teaching methodologies that bridge theoretical knowledge with practical application. Studies have shown that interactive learning environments significantly enhance students' understanding of complex security frameworks and regulatory requirements \cite{marsa2013design}. Building on this, recent investigations into the role of AI in developing cybersecurity policies highlighted both the benefits and limitations of AI-generated security documentation in educational contexts \cite{anandita2023role}.

The challenges of integrating emerging technologies into cybersecurity curricula have been well-documented. Analysis of the pedagogical implications of AI adoption in security education emphasised the need for structured guidance and critical evaluation frameworks \cite{shchavinsky2023application}. This aligns with research identifying potential risks of over-reliance on AI tools and proposing strategies for maintaining academic rigour while leveraging AI capabilities \cite{zhai2024effects}.

Recent studies have also examined the effectiveness of case-based learning in cybersecurity education \cite{anderson2024case}. Research has demonstrated how real-world scenarios and industry engagement enhance students' practical understanding of security concepts, particularly emphasising the importance of regulatory compliance and risk management in cybersecurity training \cite{alnajim2023exploring}. This aspect forms a foundation for our study's AI-assisted learning approaches.
The role of critical thinking in cybersecurity education has been extensively studied \cite{nowduri2018critical}, with proposed frameworks for evaluating student engagement with security concepts and regulatory requirements. These findings suggest that structured analytical exercises significantly improve students' ability to assess and mitigate security risks \cite{crabb2024critical}, a principle we incorporate through AI-assisted policy development and refinement.
Concerning regulatory compliance in cybersecurity education, research has examined how educational programs can better prepare students for industry requirements. Studies highlighted the importance of incorporating frameworks such as NIST and GDPR into practical exercises \cite{hajny2021framework}, an approach we extend through AI-assisted learning activities.

While existing literature demonstrates the potential of AI in education \cite{10837597,10837608,10837679}, there remains a gap in understanding how GenAI tools can be effectively integrated into cybersecurity curricula while maintaining educational integrity and ensuring regulatory compliance \cite{sandu2024role}. This study addresses this gap by presenting a structured framework for embedding AI tools within cybersecurity education, emphasising critical evaluation, research-based refinement, and practical application of security concepts.
Based on the literature review and identified research gaps, this study addresses the following research questions:
\begin{itemize}
    \item RQ1: How can GenAI tools be effectively integrated into cybersecurity education to enhance students' critical thinking and practical application skills?
    \item RQ2: What are the key challenges and success factors in implementing AI-assisted learning in cybersecurity education?
    \item RQ3: How does the integration of GenAI tools impact students' understanding of regulatory compliance and security policy development?
\end{itemize}

%%%%%%%%%%%%%%%%%%%%%%%%%%%%%%%%%%%%%%%%%%

\section{Integration Strategy and Instructional Design}
The implementation of GenAI tools in cybersecurity education required a carefully structured approach to ensure effective learning outcomes and meaningful student engagement. This section describes our pedagogical framework, implementation strategy, and specific case studies that formed the foundation of our AI-assisted learning approach. The integration strategy was designed to address the research questions outlined in Section 2, with particular emphasis on developing critical thinking skills and practical application of cybersecurity concepts.
\subsection{Pedagogical Framework}

The integration of Generative AI (GenAI) tools into cybersecurity education was designed based on constructivist learning principles \cite{olsen1999constructivist}, emphasising active engagement, critical thinking, and problem-solving. These principles position students as active participants in the learning process, where AI functions as a scaffold for foundational learning, enabling them to explore, critique, and refine AI-generated outputs within structured instructional activities. Drawing from Bloom’s Taxonomy \cite{chandio2016bloom}, this framework fosters higher-order cognitive skills such as analysis, evaluation, and synthesis, ensuring that students interact meaningfully with AI-generated content rather than consuming it passively.

Additionally, situated learning theory \cite{binti2018situated} plays a critical role in shaping this strategy. This theory suggests that knowledge is best acquired through authentic, real-world contexts, making cybersecurity an ideal domain for AI-assisted education. By embedding AI-driven tasks into cybersecurity exercises, students apply theoretical knowledge while developing practical expertise in risk assessment, policy development, and regulatory compliance. This integration encourages iterative refinement, where students critically assess AI-generated content and enhance its accuracy, alignment with regulations, and contextual applicability.

This pedagogical framework aligns with the curriculum standards of Information Security Management at the postgraduate level and  Cybersecurity Management) at the undergraduate level at the "University name anonymised" . Both courses emphasise managerial, technical, and policy-oriented skills, preparing students to tackle contemporary cybersecurity challenges. While the undergraduate course focuses on foundational cybersecurity concepts and applied exercises, the postgraduate version integrates a stronger research component, requiring students to conduct deeper analytical evaluations of AI-generated outputs. The integration of GenAI in both courses adheres to Australian higher education standards, ensuring academic rigor, industry relevance, and a strong emphasis on ethical considerations.

The strategic framework adopted for this integration is visualised in Figure 1, which outlines the pedagogical foundation and its alignment with learning objectives. The diagram highlights the two primary stages: (1) tutorial-based iterative learning and (2) assessment-driven real-world application which emphasises the role of constructivist learning principles in developing critical thinking and contextual understanding.

\begin{figure}[htbp]
    \makebox[\textwidth]{%
        \hspace*{-3cm}
        \includegraphics[width=1\textwidth]{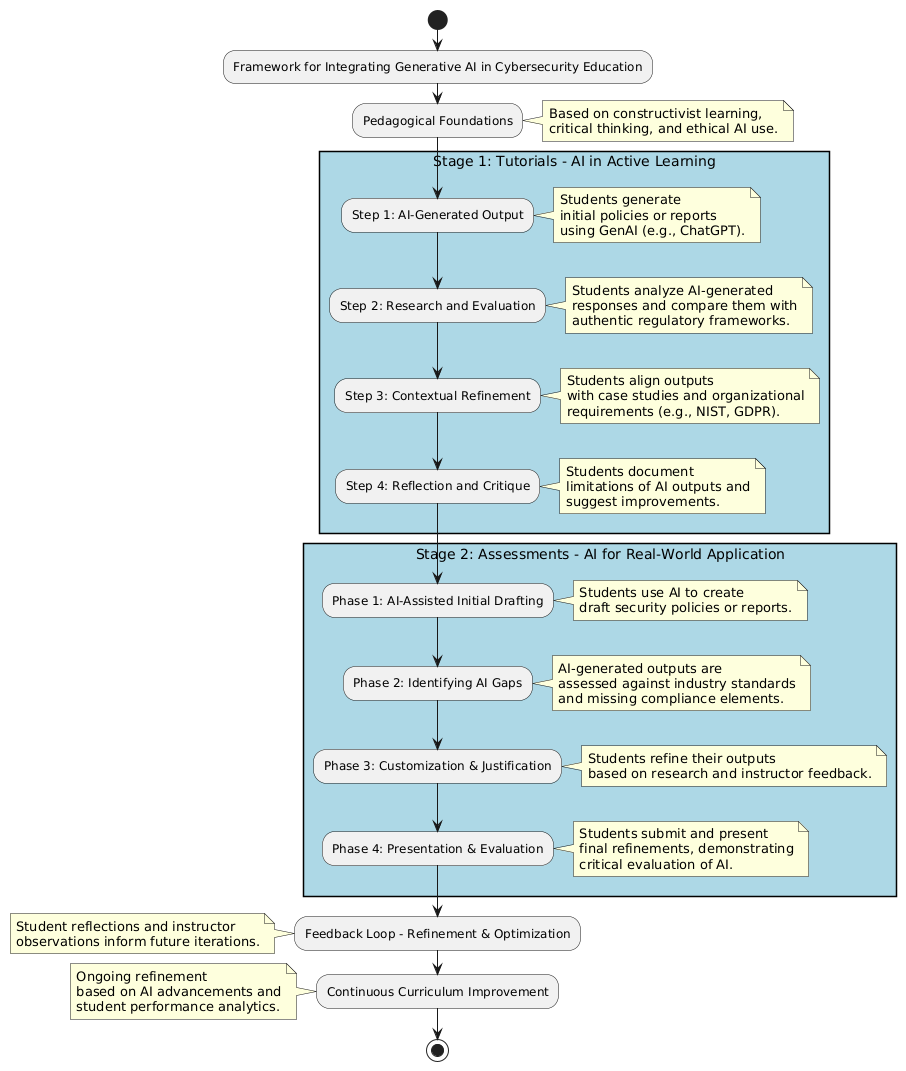}%
    }
    \caption{Strategic Framework for Integrating GenAI in Cybersecurity Education.}
    
    \label{fig:framework}
\end{figure}

\subsection{Implementation Approach}
The selection of Generative AI (GenAI) tools for cybersecurity education was guided by their versatility, accessibility, and ability to generate structured security content. ChatGPT, Gemini, and Claude were used for their capabilities in policy drafting, risk assessment, and cybersecurity strategy development. More importantly, they offered free versions.  While DeepSeek was introduced later in the term after its launch. These tools enabled students to iteratively explore cybersecurity solutions, critically assess AI-generated content, and refine outputs to align with industry standards and regulatory frameworks. However, it is noted that the majority of students relied on ChatGPT and later DeepSeek when it became available.

The implementation strategy followed a two-stage approach, embedding GenAI into both tutorials and assessments as shown in Figure 1. The first stage focused on tutorial-based iterative learning, where students engaged in structured exercises designed to foster critical engagement with AI-generated content. Tutorials followed a structured methodology, guiding students through AI-assisted content creation, research-based refinement, contextual application, and structured reflection. This iterative process ensured that students critically assessed AI-generated outputs while refining their cybersecurity strategies using academic research and regulatory frameworks.

The second stage involved assessment-driven real-world application, requiring students to apply GenAI-assisted learning to practical cybersecurity scenarios. Assessments required students to generate initial security policies or risk assessments using AI, critically evaluate gaps, and iteratively refine their work to ensure compliance with industry standards and regulatory requirements. By engaging with real-world cybersecurity case studies, students developed a nuanced understanding of AI’s role in cybersecurity management, balancing its potential benefits with its inherent limitations.
The structured process by which students engaged with GenAI tools throughout their coursework is illustrated in Figure 2. This diagram outlines the stepwise learning phases, from AI-generated output creation to research-based evaluation, contextual refinement, finalisation, and reflective analysis. The approach ensures that students move beyond passive reliance on AI-generated outputs, instead engaging deeply with cybersecurity principles an regulatory considerations.
\begin{figure}
    \centering
    \includegraphics[width=0.8\linewidth]{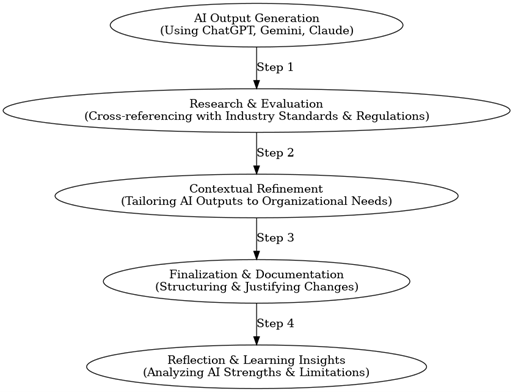}
    \caption{Stepwise AI-Driven Learning Model for Cybersecurity Education}
    \label{fig:enter-label}
\end{figure}

\subsubsection{Tutorial and Assessment Integration}
The integration strategy was structured to ensure that students progressively engaged in AI-assisted cybersecurity problem-solving. Figure 2 details the step-wise process while Table 1 summarises this approach, highlighting the distinction between tutorial-based learning and assessment-driven real-world applications, detailing instructional objectives, structured approaches, examples of exercises, and evaluation metrics.

The practical implementation of these AI-assisted learning strategies was embedded within three key case studies, each addressing a different aspect of cybersecurity education:
\begin{itemize}
    \item Security Systems Development Life Cycle (SecSDLC) in Cybersecurity Planning – Focusing on AI-generated security frameworks and their refinement.
    \item Email Policy Development for XYZ Bank – Emphasising AI’s role in policy generation and research-based enhancement.
    \item Layered Security Strategy for Financial Institutions – Highlighting the importance of prompt engineering and human oversight.
\end{itemize}
Each case study was designed to develop students' ability to analyse, refine, and apply AI-generated cybersecurity strategies in both structured tutorials and assessments.

\begin{longtable}{p{2cm} p{6.25cm} p{6.25cm}}
\caption{Comparison of Tutorials and Assessments in the Context of AI and Cybersecurity} 
\label{tab:comparison} \\

\toprule
\textbf{Aspect} & \textbf{Tutorials} & \textbf{Assessments} \\
\midrule
\endfirsthead

\multicolumn{3}{c}{{\textit{Continued from previous page}}} \\
\toprule
\textbf{Aspect} & \textbf{Tutorials} & \textbf{Assessments} \\
\midrule
\endhead

\midrule
\multicolumn{3}{r}{{\textit{Continued on next page}}} \\
\bottomrule
\endfoot

\bottomrule
\endlastfoot

\textbf{Objective} & 
Enable students to critically evaluate and refine AI-generated outputs in the context of cybersecurity. & 
Develop real-world solutions using GenAI, emphasising iterative refinement, critical analysis, and alignment with regulatory frameworks. \\

\textbf{Approach} & 
\begin{enumerate}
    \item Generate AI outputs for specific tasks.
    \item Identify gaps using academic research, industry standards, or case studies.
    \item Refine outputs to meet organisational and contextual needs.
    \item Reflect on process outcomes and learning insights.
\end{enumerate} & 
\begin{enumerate}
    \item Use GenAI to draft initial policies or solutions.
    \item Critically analyse gaps based on gathered evidence (e.g., interviews or case study data).
    \item Refine outputs to meet industry, regulatory, and organisational requirements.
    \item Present final outputs and justify modifications.
\end{enumerate} \\

\textbf{Examples} & 
\begin{itemize}
    \item \textbf{Email Policy:} Students generated an email policy for XYZ Bank using GenAI, refined it with insights from a research paper, and tailored it to align with Australian regulations.
    \item \textbf{Layered Security Strategy:} Developed a layered security strategy for a financial organisation by comparing ChatGPT’s output with requirements of the Privacy Act 1988.
    \item \textbf{NIST CSF Alignment:} Used AI-generated content to align with the NIST Cybersecurity Framework, enhancing asset management and response planning categories.
\end{itemize} & 
\begin{itemize}
    \item \textbf{Option 1 (Business Engagement):} Students interviewed a local clinic to understand security needs, used GenAI to draft a policy, and tailored it based on findings (e.g., adding wearable device management protocols).
    \item \textbf{Option 2 (Koala Health Case Study):} Students used ChatGPT to draft a security policy, identified gaps (e.g., lack of third-party accountability), and refined it to include specific protocols for pathology and pharmacy systems.
\end{itemize} \\

\textbf{Evaluation Metrics} & 
\begin{itemize}
    \item Depth of critique in identifying gaps in AI outputs (e.g., missing security controls, inadequate regulatory compliance).
    \item Quality of refinements made, ensuring alignment with research and organisational needs.
    \item Clarity and coherence of reflections on the process and GenAI’s role.
\end{itemize} & 
\begin{itemize}
    \item Completeness and feasibility of tailored refinements.
    \item Justification of modifications using evidence from interviews or case studies.
    \item Presentation clarity, including AI-generated output, gap analysis, and final recommendations.
    \item Reflection on GenAI’s strengths, limitations, and role in cybersecurity problem-solving.
\end{itemize} \\

\end{longtable}

\subsubsection{Case Study 1: Security Systems Development Life Cycle (SecSDLC) in Cybersecurity Planning}

A fundamental component of cybersecurity education involves integrating security considerations into IT system development. This tutorial introduced students to the SecSDLC model, guiding them in developing security strategies that align with cybersecurity best practices and regulatory requirements.

Students were provided with a case study scenario that required them to develop an information security plan for a financial institution. They generated an initial strategy using GenAI, which provided a structured but incomplete SecSDLC framework. The AI-generated content lacked specific compliance considerations, organisational risk factors, and threat modeling.

\begin{tcolorbox}[title=Tutorial Question on SecSDLC Implementation,
    colback=gray!5,
    colframe=gray!75!black,
    fonttitle=\bfseries]
Using the Security Systems Development Life Cycle (SecSDLC) model, develop an information security strategy for a financial institution implementing a new digital banking platform. Use GenAI to generate an initial response, then refine the AI-generated output based on regulatory requirements, risk management considerations, and corporate security objectives.
\end{tcolorbox}

The iterative refinement process engaged students in a comprehensive evaluation of AI-generated content, requiring them to critically assess the adequacy of security frameworks, risk management strategies, and regulatory compliance considerations. Throughout this process, students identified missing elements, particularly in risk assessment mechanisms and industry-specific security protocols, recognising that while AI-generated content provided a structured baseline, it often lacked depth in addressing real-world cybersecurity constraints. To enhance the applicability of AI-assisted outputs, students integrated established security frameworks, such as the NIST Cybersecurity Framework, GDPR regulations, and ISO 27001 compliance strategies \cite{singh2021review}, ensuring that their refined security plans aligned with recognised industry standards.

Beyond technical enhancements, the activity also emphasised the necessity of critical reflection on AI’s role in cybersecurity planning. Students analysed the extent to which AI-generated recommendations were applicable in practical scenarios, identifying areas where human oversight was essential in addressing contextual cybersecurity challenges. This structured engagement with AI outputs underscored the importance of expert intervention in refining cybersecurity strategies, reinforcing that while GenAI tools offer valuable starting points for security planning, they remain insufficient in capturing the full complexity of regulatory compliance and industry-specific risk considerations.

While the SecSDLC exercise focused on structuring security planning through AI-assisted frameworks, a critical component of cybersecurity management involves the creation of robust security policies. Given the increasing threat landscape and the role of human behavior in security vulnerabilities, policy development remains a key area where AI-generated content must be critically evaluated and refined. To further explore this, the next tutorial exercise guided students in leveraging AI to draft and enhance an email security policy, emphasising regulatory compliance and phishing prevention strategies.

\subsubsection{Case Study 2: Email Policy Development for XYZ Bank}

The second tutorial emphasised policy development, particularly focusing on email security policies for financial institutions. Students first generated an AI-based policy, which covered basic security controls such as encryption and password protection. However, the AI-generated policy failed to align with industry-specific compliance requirements and lacked structured phishing prevention mechanisms.

\begin{tcolorbox}[title=Tutorial Question on AI-Generated Security Policy Development,
    colback=gray!5,
    colframe=gray!75!black,
    fonttitle=\bfseries]
XYZ Bank is expanding its digital services, raising concerns about email-based cyber threats, phishing, and data breaches. Use GenAI to generate an email security policy, then refine the policy using insights from the research paper ``Analysis of the Human Factor in Cybersecurity: Identifying and Preventing Social Engineering Attacks in Financial Institutions.'' Ensure that the final version aligns with financial security regulations and best practices.
\end{tcolorbox}

To refine the AI-generated policy, students engaged in a structured evaluation of its effectiveness, focusing on the identification of security gaps and the assessment of its alignment with industry standards. Through critical analysis, they pinpointed deficiencies, such as missing compliance elements, insufficient security training measures, and inadequately defined incident response protocols, recognising that AI-generated policies often lacked the depth required for real-world applicability. Addressing these shortcomings, students integrated research-driven security enhancements, incorporating SPF, DKIM, and DMARC authentication mechanisms to strengthen email security and mitigate phishing risks.

Beyond technical refinements, students ensured that their revised policies aligned with financial sector regulations, incorporating compliance measures that adhered to APRA guidelines and broader cybersecurity frameworks governing the financial industry. The process demonstrated that while AI can generate a preliminary policy framework, it lacks the specificity required for regulatory compliance and organisational security needs. This case study reinforced the importance of research-driven refinement, illustrating that AI-generated security policies must be critically assessed and enhanced to meet established industry standards and effectively address sector-specific cybersecurity challenges.

In addition to security policy development, cybersecurity management requires a more comprehensive approach that integrates multiple layers of protection across an organisation's infrastructure. Recognising that AI-generated policies often provide a static, one-dimensional perspective on security, the following tutorial expanded the scope by engaging students in designing a multi-layered security strategy for financial institutions. This exercise reinforced the importance of prompt engineering in AI-generated security recommendations while demonstrating that, even with a well-crafted prompt, human oversight remains essential in aligning security measures with industry best practices and regulatory mandates.

\subsubsection{Case Study 3: Layered Security Strategy for Financial Institutions}

A key tutorial exercise involved developing a layered security strategy for a financial institution. Students used ChatGPT to generate an initial security framework, which included physical security, network protection, and incident response. However, upon further analysis, students identified missing provisions for cross-border data encryption and inadequate access control mechanisms. Through research-based refinement, students integrated NIST Cybersecurity Framework standards, encryption protocols, and multi-factor authentication mechanisms, ensuring that the final strategy adhered to sector-specific security regulations.

Similarly, in assessments, students were required to apply AI-generated cybersecurity policies to real-world case studies. In one assessment, students collaborated with local businesses and industry partners to develop a customised cybersecurity policy based on real-world organisational needs. Another assessment required students to analyse the Koala Health case study, critically evaluating AI-generated policies to address cybersecurity challenges related to telehealth services and wearable device management. 

The assessment design ensured that students not only identified security gaps in AI-generated outputs but also justified their refinements using established industry standards. Furthermore, by critically reflecting on AI’s role in cybersecurity decision-making, students gained a deeper appreciation of both the strengths and limitations of AI-assisted learning.
Table 2 provides the key learning objectives and examples of GenAI Use in Tutorials and Assessments.

\begin{table}[htbp]
\centering
\caption{Key Learning Objectives and Examples of GenAI Use in Tutorials and Assessments}
\label{tab:learning_objectives}
\begin{tabular}{p{0.3\textwidth} p{0.3\textwidth} p{0.3\textwidth}}
\toprule
\textbf{Learning Objective} & \textbf{Tutorial Example} & \textbf{Assessment Example} \\
\midrule
Develop critical thinking by identifying gaps in AI outputs. & Critiquing ChatGPT’s output for an email policy to identify missing compliance measures. & Identifying gaps in a security policy drafted for a small business or Koala Health. \\
\midrule
Apply industry standards to AI-generated content. & Comparing AI-generated security strategies with the NIST Cybersecurity Framework. & Aligning AI-generated policies with the Australian Privacy Act 1988. \\
\midrule
Tailor solutions to specific organisational contexts. & Refining a layered security strategy with regional compliance considerations. & Customising a policy to a local business’s unique security needs based on interviews. \\
\midrule
Reflect on the strengths and limitations of GenAI in practice. & Documenting insights into AI capabilities and areas requiring human expertise. & Reflecting on the role of GenAI in crafting security policies and compliance measures. \\
\bottomrule
\end{tabular}
\end{table}

By engaging students in structured AI-driven exercises, ranging from security planning (SecSDLC) to policy refinement and layered security strategies, this study demonstrated how GenAI can be integrated into cybersecurity education to enhance critical thinking and analytical skills. However, as these case studies illustrate, AI-generated outputs are only as effective as the critical engagement they receive from students and educators. Therefore, by embedding AI tools within structured instructional activities, this approach fosters a balanced learning environment where AI serves as an enabler of critical thinking rather than a substitute for human expertise. The next section presents the results of this approach, examining its impact on student learning, engagement, and the refinement of cybersecurity knowledge.

\section{Results}
The results of integrating GenAI into cybersecurity education are examined through student feedback, instructor observations, and an analysis of the challenges and unexpected outcomes encountered during implementation. These insights highlight the effectiveness of AI-assisted learning, while also addressing areas requiring further refinement.

\subsection{Student Feedback and Learning Reflections}
Student reflections provided valuable insights into the reception and effectiveness of GenAI integration within the curriculum. Overall, students responded positively to the structured approach, particularly appreciating the role of AI in streamlining the initial drafting process. Many noted that GenAI served as a useful starting point, enabling them to focus on refining outputs rather than generating content from scratch. One student remarked:

\begin{quote}
\textit{``The AI-generated policy was a great starting point, but refining it based on research allowed me to understand the gaps and create a more tailored solution.''}
\end{quote}

Similarly, another student emphasised the efficiency gains:

\begin{quote}
\textit{``Using GenAI helped me focus on improving content rather than starting from scratch, saving time and allowing me to dive deeper into compliance and security details.''}
\end{quote}

Beyond efficiency, student feedback underscored the importance of human oversight in refining AI-generated outputs. Several students reflected on the realisation that AI tools often produce generic or incomplete responses that require critical evaluation and domain-specific expertise. This iterative refinement process contributed to a deeper engagement with regulatory frameworks and industry standards, reinforcing the necessity of balancing automation with analytical reasoning.

\subsection{Instructor Observations on Learning Outcomes}
From an instructional perspective, the structured integration of GenAI significantly enhanced student engagement, critical thinking, and contextual application of cybersecurity concepts. This phased approach where students first generated AI-based content, then critically evaluated and refined it, led to higher levels of analytical reasoning compared to traditional instructional methods.

Instructors observed a notable improvement in students' ability to assess regulatory compliance and adapt security policies to organisational needs. The use of real-world scenarios helped bridge the gap between theoretical knowledge and practical cybersecurity applications, fostering a more holistic understanding of risk management, compliance requirements, and policy development. While some students initially struggled with evaluating AI-generated content, reflection activities and structured research prompts played a critical role in deepening their analytical engagement.

\subsection{Challenges Encountered During Implementation}
Despite its overall effectiveness, the integration of GenAI within the curriculum presented several pedagogical and logistical challenges that impacted both student learning and instructional design.

A primary challenge was the over-reliance on AI-generated outputs among some students. Certain students treated AI-generated responses as authoritative rather than as an initial draft requiring further refinement. This was particularly evident in tasks such as policy development, where students initially submitted responses that lacked depth or contextual alignment with cybersecurity regulations. For instance, in exercises involving email policy refinement, students struggled to identify jurisdiction-specific compliance requirements, such as those outlined in the Privacy Act 1988. Addressing this challenge required additional scaffolding through explicit prompts and guided discussions, emphasising the limitations of AI-generated content and the need for rigorous evaluation and adaptation.

Another challenge was the variability in students' familiarity with AI tools. While some students demonstrated proficiency in leveraging GenAI for cybersecurity policy drafting and security strategy development, others required substantial guidance in interpreting AI outputs and identifying relevant modifications. This skill disparity led to an uneven starting point for the cohort, requiring instructors to provide targeted support for students less experienced in evaluating AI-assisted cybersecurity solutions.

A further difficulty arose in context-specific customisation of AI outputs. AI-generated security frameworks and policies often lacked the specificity required to align with organisational contexts and regulatory mandates. In assessments requiring students to draft security policies for real businesses, aligning AI-generated outputs with industry-specific standards such as the Australian Privacy Act and GDPR proved to be a complex and time-intensive task. Students had to engage in extensive research and iterative refinement to ensure their policies met both operational and regulatory requirements, reinforcing the necessity of domain-specific expertise.

Also, assessments that required real-world business engagement posed logistical constraints. While students benefited from working with industry partners, coordinating interviews with business stakeholders presented scheduling difficulties. This challenge was partially addressed by providing a case study alternative, allowing students to apply their knowledge to a controlled scenario rather than relying solely on external collaborations. However, this issue highlighted the inherent complexities of integrating authentic, real-world interactions within structured academic settings.

\subsection{Unexpected Benefits and Drawbacks}
Beyond its intended learning objectives, the integration of GenAI produced several unexpected benefits and challenges that further shaped the instructional strategy.

A significant advantage was the increased engagement and motivation among students, particularly in assignments that simulated real-world cybersecurity challenges. The iterative process of drafting, refining, and presenting AI-assisted security policies encouraged students to take ownership of their work, fostering a more immersive and dynamic learning environment. The interactive nature of AI tools also promoted creativity, allowing students to explore multiple policy variations before finalising their submissions.

Another key benefit was the reinforcement of regulatory knowledge through hands-on refinement of AI-generated security strategies. For instance, students tasked with improving AI-generated layered security strategies for financial institutions gained practical exposure to encryption protocols for cross-border data transfers, incident response planning, and compliance measures tailored to Australian cybersecurity regulations. This process deepened their understanding of the interplay between cybersecurity frameworks and real-world implementation.

However, the integration of AI also presented certain drawbacks, particularly in balancing cognitive load and task complexity. Some students felt overwhelmed by the dual challenge of critiquing AI-generated outputs while simultaneously aligning them with regulatory standards. The iterative nature of the exercises, while beneficial for deep learning, also increased the time required to complete assignments, occasionally leading to frustration among students with weaker research or analytical skills. Instructors had to provide additional support and structured guidance to ensure that students were able to navigate AI outputs effectively without becoming disengaged.

One of the most valuable takeaways for students was the realisation of AI's inherent limitations. Many observed that while AI-assisted learning facilitated faster drafting and content structuring, it frequently lacked contextual specificity, regulatory depth, and enforcement mechanisms. As one student noted:

\begin{quote}
\textit{``The AI-generated policy was a good outline, but it lacked enforcement mechanisms and did not include risk mitigation plans for third-party vendors. AI alone is insufficient for developing robust cybersecurity policies.''}
\end{quote}

This reflection reinforced the broader lesson that AI tools must be viewed as assistive rather than authoritative, requiring human expertise to refine and contextualise outputs.

Additionally, Table 3 provides further insights into the instructional impact of AI integration. It summarises key student reflections, instructor observations, challenges, and emerging trends that shaped the overall learning experience.

\begin{longtable}{p{3cm} p{6cm} p{6cm}}
\caption{Comparative Analysis of Tutorials and Assessments} \label{tab:comparative_analysis} \\
\toprule
\textbf{Aspect} & \textbf{Tutorials} & \textbf{Assessments} \\
\midrule
\endfirsthead

\multicolumn{3}{c}{{\textit{Continued from previous page}}} \\
\toprule
\textbf{Aspect} & \textbf{Tutorials} & \textbf{Assessments} \\
\midrule
\endhead

\midrule
\multicolumn{3}{r}{{\textit{Continued on next page}}} \\
\bottomrule
\endfoot

\bottomrule
\endlastfoot

\textbf{Student Feedback} & 
\begin{itemize}
    \item AI provided a good starting point but lacked depth; tailoring it was a valuable learning experience.
    \item Reflection steps helped me see gaps in AI output and how academic research fills those gaps.
\end{itemize} & 
\begin{itemize}
    \item Using GenAI to draft policies saved time but required significant adjustments to meet regulatory needs.
    \item The engagement with real-world scenarios made the task more meaningful.
\end{itemize} \\

\textbf{Instructor Observations} & 
\begin{itemize}
    \item Students struggled initially with evaluating AI outputs but improved after guided research tasks.
    \item Reflection activities enhanced critical thinking and contextual application skills.
\end{itemize} & 
\begin{itemize}
    \item Students showed higher engagement when interacting with local businesses.
    \item Refinements demonstrated a clear understanding of regulatory and organisational contexts.
\end{itemize} \\

\textbf{Challenges Encountered} & 
\begin{itemize}
    \item Some students relied too heavily on AI without adequately critiquing outputs.
    \item Identifying gaps required substantial scaffolding and prompts.
\end{itemize} & 
\begin{itemize}
    \item Real-world engagements posed logistical challenges (e.g., scheduling interviews with business stakeholders).
    \item Students with weaker research skills struggled to justify refinements.
\end{itemize} \\

\textbf{Unexpected Benefits} & 
\begin{itemize}
    \item AI tools encouraged creativity, particularly in refining policies.
    \item Students became more confident in using AI as a collaborative tool rather than a replacement for expertise.
\end{itemize} & 
\begin{itemize}
    \item Engagement with businesses enhanced students' professional communication skills.
    \item Case study options provided a controlled environment for students who struggled with interviews.
\end{itemize} \\

\textbf{Drawbacks} & 
\begin{itemize}
    \item AI outputs sometimes introduced errors or irrelevant content, requiring careful analysis to avoid misinformation.
\end{itemize} & 
\begin{itemize}
    \item Some students felt overwhelmed by the dual task of critiquing AI outputs and aligning them with regulatory frameworks.
\end{itemize} \\

\textbf{Insights for Future} & 
\begin{itemize}
    \item Provide additional guidance on critiquing AI outputs and identifying gaps.
    \item Develop more targeted tutorials to address specific weaknesses (e.g., regulatory compliance).
\end{itemize} & 
\begin{itemize}
    \item Simplify business engagement logistics by offering predefined templates for interviews.
    \item Include examples of refined outputs to clarify expectations.
\end{itemize} \\

\end{longtable}

\section{Discussion:}
The integration of Generative AI (GenAI) into cybersecurity education has demonstrated significant pedagogical value, enhancing critical thinking, contextual application, and problem-solving abilities. This section discusses the effectiveness of this instructional strategy in meeting learning objectives, compares it to traditional teaching approaches, identifies best practices, and provides recommendations for improvement. Furthermore, it examines the broader implications of GenAI-assisted learning in cybersecurity education.

\subsection{Effectiveness in Meeting Learning Objectives}
The structured use of GenAI tools successfully fostered critical evaluation, research-driven refinement, and real-world application of cybersecurity principles. Student feedback and performance data indicate that AI-assisted learning improved problem-solving capabilities and regulatory awareness, while also encouraging deeper engagement with cybersecurity frameworks.
Figure 1 highlights key student reflections, demonstrating that most students valued AI as a starting point for drafting cybersecurity policies and security strategies, but also acknowledged the need for further refinement and contextualisation. A notable proportion of students reported challenges in identifying gaps in AI-generated content, which underscores the necessity of structured research-based refinement exercises.
\begin{figure}
    \centering
    \includegraphics[width=1\linewidth]{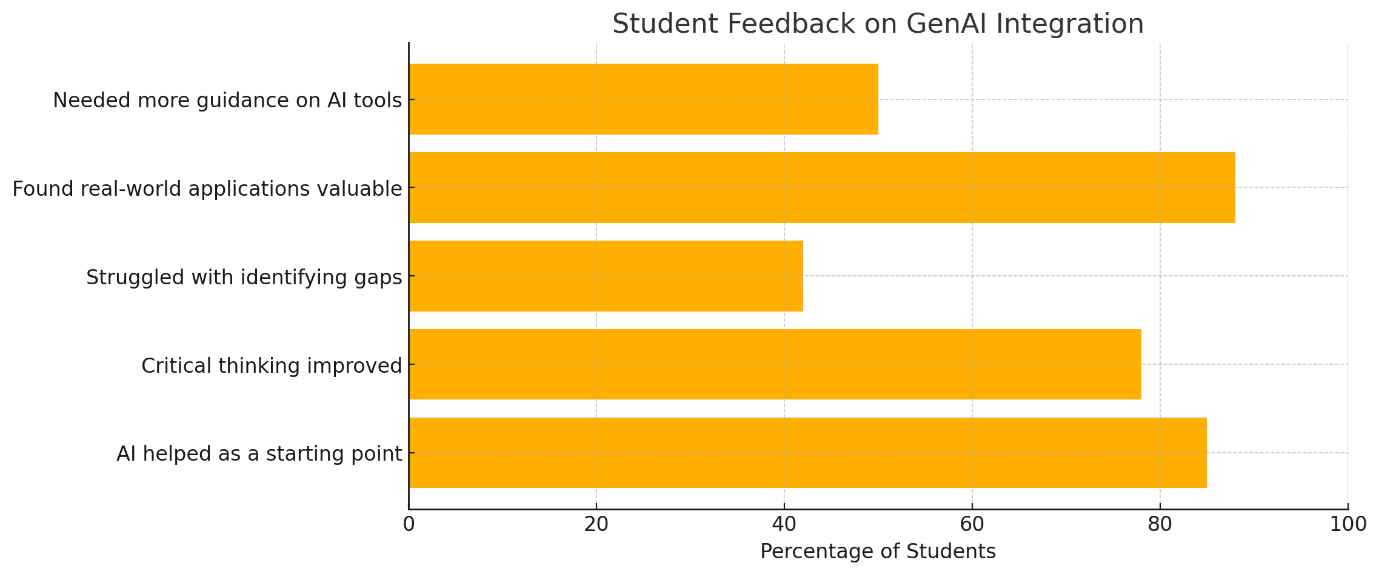}
    \caption{Students Feedback}
    \label{fig:enter-label}
\end{figure}
The effectiveness of different instructional strategies in promoting learning outcomes is further illustrated in Figures 3 and 4, which map teaching strategies to specific learning objectives. Notably, assessments and case study-based exercises had the highest impact on regulatory awareness and problem-solving abilities, while tutorials played a critical role in improving AI confidence and research skills.

\begin{figure}
    \centering
    \includegraphics[width=1\linewidth]{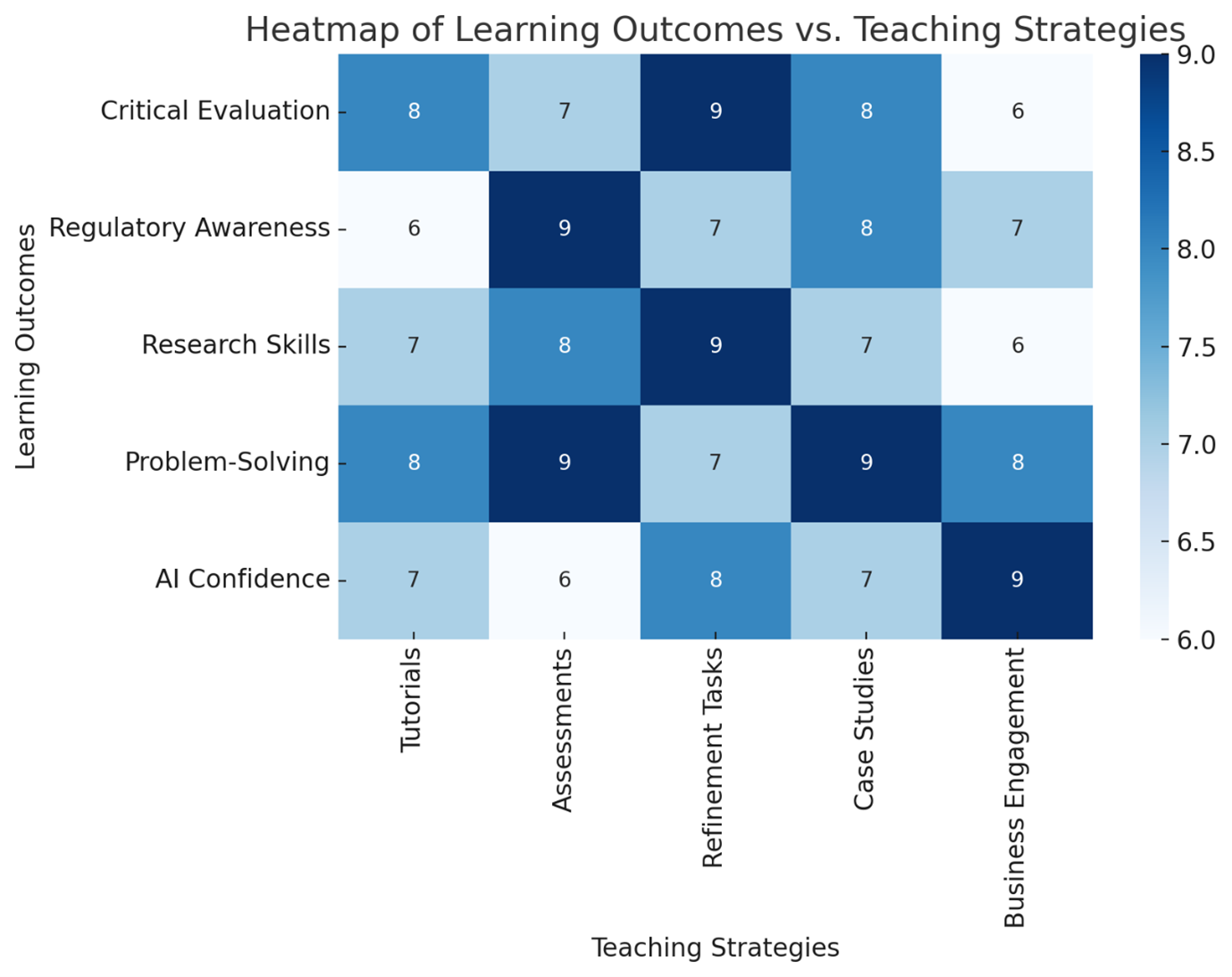}
    \caption{Heatmap}
    \label{fig:enter-label}
\end{figure}
Table 4 further substantiates these findings, demonstrating how the integration of AI tools across tutorials and assessments facilitated real-world problem-solving, alignment with regulatory standards, and critical evaluation of AI-generated security policies.

\begin{longtable}{>{\raggedright\arraybackslash}p{3cm} >{\raggedright\arraybackslash}p{6cm} >{\raggedright\arraybackslash}p{5.5cm}}
\caption{Mapping Student Learning Outcomes to Teaching Strategies}
\label{tab:learning_outcomes} \\
\toprule
\textbf{Learning Outcome} & \textbf{Tutorials (Guided Exercises)} & \textbf{Assessments (Real-World Engagement)} \\
\midrule
\endfirsthead
\multicolumn{3}{c}%
{{\bfseries \tablename\ \thetable{} -- continued from previous page}} \\
\toprule
\textbf{Learning Outcome} & \textbf{Tutorials (Guided Exercises)} & \textbf{Assessments (Real-World Engagement)} \\
\midrule
\endhead
\bottomrule
\multicolumn{3}{r}{{Continued on next page}} \\
\endfoot
\bottomrule
\endlastfoot

Critical Evaluation of AI Outputs & 
Students analysed AI-generated email policies, identified gaps, and refined them using academic research. & 
Students compared AI-generated security policies with industry best practices and regulatory requirements. \\
\midrule

Regulatory Awareness & 
Exercises required students to align AI outputs with NIST, GDPR, and Australian regulations. & 
Business engagement and case studies exposed students to industry-specific compliance challenges. \\
\midrule

Application of Research Skills & 
Students reviewed research papers to validate AI-generated content. & 
Assessments required integrating insights from stakeholder interviews or real-world case studies. \\
\midrule

Real-World Problem Solving & 
AI-generated layered security strategies were enhanced using jurisdiction-specific security frameworks. & 
Students developed refined policies that addressed organisational needs, including BYOD and third-party security risks. \\
\midrule

Confidence in AI as a Tool & 
Students gained awareness of AI's limitations and learned to improve upon its outputs. & 
The iterative process demonstrated the necessity of human oversight, reinforcing AI's role as an assistive tool rather than a replacement. \\
\end{longtable}

\subsection{Comparison with Traditional Teaching Approaches}
Compared to conventional lecture-based methods, the integration of AI-assisted learning strategies provided a more interactive, hands-on learning experience. Traditional approaches often rely on passive knowledge acquisition, where students primarily engage with cybersecurity frameworks through lectures and reading materials. In contrast, the structured AI-driven exercises encouraged active participation, requiring students to generate, critique, and refine AI outputs using industry standards and academic research.
The iterative nature of the exercises bridged the gap between theoretical knowledge and practical application, a limitation often observed in traditional cybersecurity education. The business engagement component further exposed students to real-world cybersecurity challenges, fostering a deeper understanding of risk management, compliance, and security policy adaptation. These findings align with broader trends in cybersecurity education, emphasising experiential learning as a crucial factor in preparing students for industry challenges.
However, the introduction of AI into cybersecurity education also introduced unique challenges, particularly regarding student reliance on AI-generated content and variability in AI literacy levels. As seen in Table 5, a significant portion of students initially struggled to critically evaluate AI-generated outputs, indicating that additional scaffolding and guidance may be required to ensure a balanced use of AI as a learning tool.

\subsection{Best Practices, Future Recommendations, and Broader Implications}

The findings from this study highlight key best practices for integrating GenAI into cybersecurity education, emphasising structured AI engagement, external research integration, and real-world application. While the implementation of GenAI demonstrated pedagogical benefits, several areas for improvement emerged, necessitating refinements to instructional strategies and broader considerations for the future of AI in cybersecurity education.

\begin{table*}[t]
\caption{Challenges and Solutions in AI Integration}
\label{tab:challenges_solutions}
\centering
\small
\begin{tabularx}{\textwidth}{>{\raggedright\arraybackslash}p{5cm} >{\raggedright\arraybackslash}p{6cm}}
\toprule
\textbf{Challenge} & \textbf{Solution Implemented} \\
\midrule
Over-reliance on AI outputs & 
Incorporated structured reflection exercises to highlight AI limitations. \\
\midrule
Skill variability among students & 
Provided additional scaffolding and guided tutorials for students new to AI. \\
\midrule
Difficulties in context-specific customisation & 
Emphasised research and regulatory alignment through targeted exercises. \\
\bottomrule
\end{tabularx}
\end{table*}
\subsubsection{ Best Practices for AI-Assisted Learning in Cybersecurity}
The structured integration of GenAI into tutorials, assessments, and case-based exercises provided students with an opportunity to develop analytical skills, critically engage with AI-generated outputs, and refine cybersecurity strategies based on regulatory frameworks. Our implementation revealed several key best practices that enhanced the educational experience. The provision of structured guidance on AI tool usage proved essential, ensuring that students actively critiqued and refined AI-generated content rather than passively accepting outputs. In this context, instructor-led scaffolding played a crucial role in helping students recognise both the strengths and limitations of AI-generated security policies and risk assessments.

Critical thinking and regulatory awareness were reinforced through the incorporation of external research components, which encouraged students to validate AI-generated responses against established industry standards such as the NIST Cybersecurity Framework, GDPR, and ISO 27001. Additionally, the embedding of real-world case studies and business engagement activities helped contextualise AI-assisted learning within authentic cybersecurity challenges, fostering practical application and regulatory alignment. By integrating these best practices into cybersecurity education, educators can maximise the pedagogical benefits of AI while effectively mitigating its inherent limitations.

\subsubsection{Recommendations for Future Implementation}
While the integration of GenAI significantly enhanced student engagement and real-world problem-solving skills, challenges such as AI literacy gaps, regulatory adaptation, and student variability in AI competency necessitate targeted improvements. Table 6 summarises key recommendations based on instructor observations and student feedback, outlining actionable refinements for AI-assisted instruction.

Future implementations should address several critical areas for improvement. Our observations highlighted the need for enhanced AI literacy support, as some students required additional guidance in navigating AI-generated outputs, particularly in differentiating accurate content from AI hallucinations. To address this, introducing AI literacy modules and structured evaluation rubrics can help students develop a balanced understanding of AI's role in cybersecurity.

The alignment with jurisdiction-specific regulations emerged as another key focus area, as students faced challenges in adapting AI-generated security policies to specific compliance frameworks. To strengthen this aspect, developing targeted tutorials on regional regulations, such as the Australian Privacy Act, GDPR, and PCI DSS, will reinforce the importance of regulatory adherence in cybersecurity decision-making.
Additionally, the varying levels of student proficiency called for developing tiered exercises to accommodate different skill levels. While some students thrived in AI-assisted problem-solving, others required additional support. Implementing adaptive learning approaches, including tiered case studies and optional advanced cybersecurity scenarios, will provide differentiated learning opportunities that cater to both foundational and advanced learners.

These refinements will ensure that AI-assisted cybersecurity education remains both effective and scalable, preparing students for dynamic and evolving security challenges.

\begin{table*}[t]
\caption{AI Integration Recommendations}
\label{tab:ai_recommendations}
\begin{tabular*}{\textwidth}{p{2.8cm} p{5cm} p{5cm}}
\hline\noalign{\smallskip}
\textbf{Focus Area} & \textbf{Instructor Observation} & \textbf{Recommendations} \\
\noalign{\smallskip}\hline\noalign{\smallskip}
\textbf{AI Literacy \& Student Readiness} & 
Some students struggled with AI tools and relied too much on generated outputs without critical engagement. & 
Introduce AI literacy modules, hands-on tutorials, and mini AI-based exercises to build foundational AI understanding. \\
\noalign{\smallskip}
\textbf{Critical Thinking \& AI Output Analysis} & 
Students needed substantial guidance to assess the accuracy and relevance of AI-generated cybersecurity policies. & 
Develop structured evaluation rubrics, encourage peer-review sessions, and provide side-by-side AI-human comparisons. \\
\noalign{\smallskip}
\textbf{Contextualisation \& Regulatory Compliance} & 
AI-generated content lacked depth in regulatory compliance, requiring manual refinement to meet industry standards. & 
Require students to refine AI outputs by aligning them with industry standards (e.g., NIST, GDPR) and jurisdiction-specific regulations. \\
\noalign{\smallskip}
\textbf{Skill Variability \& Adaptive Learning} & 
There was a noticeable skill gap between students familiar with AI tools and those new to them, affecting learning outcomes. & 
Implement tiered learning approaches, offer AI-generated exemplars, and allow optional AI-assisted assessments. \\
\noalign{\smallskip}
\textbf{Ethical AI Use \& Academic Integrity} & 
Concerns arose over potential AI misuse, such as over-reliance on GenAI for assignments and automated plagiarism. & 
Establish AI usage policies, introduce self-reflection logs on AI use, and integrate AI-detection tools in grading. \\
\noalign{\smallskip}
\textbf{Real-World Engagement \& Industry Alignment} & 
Students found real-world applications highly beneficial, but logistical issues (e.g., scheduling interviews) limited engagement. & 
Create structured business engagement templates, formalise industry collaborations, and introduce 'AI in Practice' guest lectures. \\
\noalign{\smallskip}
\textbf{Optimised AI-Assisted Assessments} & 
Assessments benefited from AI assistance, but students needed to explain modifications rather than simply submitting AI-refined content. & 
Shift to iterative AI-assisted assignments with critique-refinement cycles, scenario-based AI tasks, and explanation-focused grading. \\
\noalign{\smallskip}\hline
\end{tabular*}
\end{table*}

\subsubsection{Broader Implications for Cybersecurity Education}
The integration of GenAI in cybersecurity education represents a significant shift in pedagogical strategies, reinforcing the role of AI literacy, ethical considerations, and regulatory compliance in modern cybersecurity curricula. By combining AI-assisted learning with critical evaluation, research-based refinement, and real-world application, this approach fosters a more adaptive, problem-solving-oriented educational experience.

Beyond the classroom, these findings align with emerging industry trends, where AI is increasingly leveraged for automated security policy generation, threat detection, and compliance management. However, the limitations observed in AI-generated security policies highlight the necessity of human expertise in refining and contextualising AI outputs. This underscores the importance of educating future cybersecurity professionals to use AI as an assistive tool rather than a replacement for analytical reasoning.

Looking ahead, future research should explore the long-term impact of AI-driven cybersecurity education, particularly in relation to knowledge retention, adaptability to emerging threats, and industry preparedness. Additionally, investigating the potential of adaptive AI-learning models, such as personalised AI-assisted assessments and AI-enhanced threat simulations, could provide deeper insights into how AI can be leveraged to strengthen cybersecurity training in both academic and professional contexts.

By integrating structured AI engagement, research-based refinement, and practical cybersecurity applications, this study presents a scalable framework for AI-assisted education. Moving forward, ensuring a balance between AI-driven automation and human analytical oversight will be critical in preparing cybersecurity professionals for the evolving landscape of digital security threats and regulatory requirements.

\section{ Limitations}
While this study demonstrates the potential of Generative AI (GenAI) in cybersecurity education, several limitations must be acknowledged. The findings are based on a single academic term, and the integration of AI tools within the curriculum was subject to evolving technological capabilities and student familiarity with AI-driven learning. As a result, the effectiveness of AI-assisted instruction may vary across different cohorts, institutions, and pedagogical contexts.

A key limitation pertains to the variability in students’ prior experience with AI tools, which influenced their ability to critically assess and refine AI-generated content. While some students demonstrated strong analytical skills in evaluating AI outputs against regulatory frameworks and security best practices, others exhibited a tendency to accept AI-generated responses uncritically. This disparity required additional scaffolding and instructor intervention, highlighting the challenge of ensuring consistent engagement across a diverse student population.

Moreover, while AI tools facilitated the rapid generation of cybersecurity policies and risk assessments, their outputs often lacked contextual specificity and regulatory depth. This limitation underscores the necessity of human oversight in validating AI-generated recommendations, reinforcing the idea that AI should be regarded as an assistive tool rather than a definitive authority on cybersecurity practices. Additionally, the reliance on publicly available AI models introduced concerns regarding data accuracy, potential biases, and the evolving nature of AI training datasets, which may impact the reproducibility of results in future studies.

Technical limitations of the GenAI tools used also impacted the study's implementation. The reliance on publicly available versions of ChatGPT, Claude, and DeepSeek meant that students experienced occasional service disruptions and usage limitations. These tools' responses sometimes showed inconsistency across different sessions, particularly in generating security policies and regulatory compliance documentation. Additionally, the AI models' knowledge cutoff dates meant that coverage of the newest cybersecurity threats and regulations required manual updates and instructor intervention. The inability to fine-tune these models for specific cybersecurity education needs also limited the customisation of AI-generated content for different skill levels and learning objectives.

Finally, the study was conducted within a structured academic setting where students operated under guided exercises and assessments. The extent to which these findings generalise to professional cybersecurity environments, where AI tools are employed in dynamic and high-risk decision-making contexts, remains an area for further investigation. Addressing these limitations in future research will be crucial to refining AI-assisted learning strategies and ensuring their long-term viability in cybersecurity education.

\section{Conclusion}

This study introduced a structured framework for integrating Generative AI (GenAI) into cybersecurity education, demonstrating its potential to enhance critical thinking, regulatory awareness, and practical problem-solving skills. The implementation followed a two-stage approach that embedded GenAI within tutorial exercises for iterative learning and integrated AI-assisted tasks into assessments for real-world application, providing students with comprehensive, hands-on engagement with AI-driven cybersecurity strategies. Through case studies focused on security policy development, the Security Systems Development Life Cycle (SecSDLC), and layered security strategies, the study showcased how AI can be leveraged to support learning while maintaining the rigour of cybersecurity education.

Key findings indicate that GenAI significantly streamlined initial content generation, enabling students to focus on evaluating, refining, and aligning outputs with industry standards and regulatory requirements. However, the study also highlighted challenges related to over-reliance on AI and variability in AI literacy, underscoring the need for structured guidance and research-driven refinement. The effectiveness of AI-assisted learning was contingent on three critical factors: clear instructional scaffolding, emphasis on research-informed modifications, and contextual alignment with real-world cybersecurity challenges.

By positioning AI as an assistive tool rather than a substitute for human expertise, this study reinforces the value of combining automation with analytical reasoning in cybersecurity education. Future research should explore long-term impacts on student competency, adaptive AI-learning models, and scalable implementations of AI-assisted instruction in cybersecurity training.

\backmatter

\bmhead{Supplementary information}

This article includes one sample tutorial as supplementary material to support the reproducibility of the proposed teaching strategies. The supplementary file provides a detailed step-by-step example of how Generative AI is integrated into a cybersecurity education tutorial.

\bmhead{Acknowledgements}

The authors wish to extend sincere gratitude to the students who actively participated in this study. Their engagement in tutorials, discussions, and assessments provided invaluable insights into the integration of Generative AI in cybersecurity education. Beyond formal reflections, informal feedback gathered during class discussions and Zoom meetings played a crucial role in shaping the instructor’s observations and refining the pedagogical approach. Their willingness to share their experiences, challenges, and suggestions significantly contributed to the continuous improvement of the instructional design and assessment strategies.

The authors also acknowledge the support of *anonymised University name* for providing the academic environment necessary to explore innovative teaching methodologies.

Additionally, ChatGPT was used to assist in improving the clarity, structure, and coherence of the manuscript based on original content provided by the author. While the research design, framework development, and intellectual contributions remain solely the authors' work, ChatGPT was also utilised for brainstorming titles, refining technical descriptions, and converting content into LaTeX where applicable.

\section*{Declarations}

\noindent This manuscript complies with the journal's submission requirements regarding declarations. The relevant information is provided below.

\subsection*{Funding}
This research received no specific grant from any funding agency, commercial, or not-for-profit sectors.

\subsection*{Conflict of Interest}
The authors declare that they have no competing interests.

\subsection*{Ethics Approval and Consent to Participate}
Ethical approval was not required for this study as it did not involve human participants in research requiring informed consent. The study analysed student responses to tutorial exercises that were conducted as part of regular coursework. The findings are derived from instructor observations and students’ written reflections, which were reviewed at an aggregate level without collecting identifiable data.

\subsection*{Consent for Publication}
Not applicable.

\subsection*{Data Availability}
All relevant data analysed during this study are included in the manuscript. Additional supporting materials, such as a sample tutorial, are provided as supplementary material.

\subsection*{Materials Availability}
Not applicable.

\subsection*{Code Availability}
Not applicable.

\subsection*{Authors’ Contributions}
\textbf{Mahmoud Elkhodr}: Conceptualisation, methodology, data collection, analysis, and manuscript writing.  

\textbf{Ergun Gide}: Provided mentorship, contributed to discussions, and reviewed the manuscript.  
Both authors read and approved the final manuscript.

\bigskip

\begin{appendices}

\section{Sample Tutorial: Developing an Email Policy Using Generative AI}

\subsection{Case Scenario: XYZ Bank}

XYZ is an Australian bank that offers banking, investment, and financial services to small businesses and the public. The bank has multiple branches in all state capitals in Australia and plans to expand into rural areas through a loan program.

XYZ’s organisational structure consists of four major divisions: Business, IT, Accounting and Finance, and Human Resources. Employees communicate via an internal network, and free wireless LAN access is provided to customers and visitors. The bank also offers online banking services, and customer service representatives access customer data for loan approvals, financial profiling, and marketing.

Due to a recent series of information security attacks, XYZ’s Chief Information Security Officer (CISO) is determined to improve the bank’s security systems, particularly in email security policies.

\subsection{Task: Developing an Email Policy for XYZ Bank}

The objective of this tutorial is to help students understand:
\begin{itemize}
    \item The role of \textbf{Generative AI in drafting cybersecurity policies}.
    \item How to \textbf{critically analyse AI-generated outputs}.
    \item The importance of \textbf{regulatory compliance and security best practices} in email communication.
\end{itemize}

The tutorial consists of three parts:

\subsubsection{Part 1: Generate an Initial Policy Using Generative AI}

\begin{itemize}
    \item Use a Generative AI tool (e.g., ChatGPT, Gemini, Claude) to \textbf{generate an Email Policy} for XYZ Bank based on the provided case scenario.
    \item Copy and paste the \textbf{AI-generated policy} into the submission.
    \item \textbf{Important:} Do not edit the AI-generated response at this stage.
\end{itemize}

\subsubsection{Part 2: Develop a Research-Based Policy}

\noindent Read the research paper:
\textit{“Analysis of the Human Factor in Cybersecurity: Identifying and Preventing Social Engineering Attacks in Financial Institutions”} by Ibrahim Momoh, Gabriel Adelaja, and Ghaffar Ejiwumi.

Using the key insights from the paper, revise and refine the email policy by integrating:
\begin{itemize}
    \item \textbf{Overview and Purpose:}  
    Highlight email security risks and the importance of \textbf{phishing awareness training}.
    \item \textbf{Acceptable Use and Behaviour:}  
    Establish rules prohibiting employees from clicking on unknown links or opening unsolicited attachments.
    \item \textbf{Security Controls:}  
    Implement measures such as \textbf{blocking auto-forwarding to external email addresses}, enforcing \textbf{multi-factor authentication (MFA)}, and \textbf{email encryption for sensitive communication}.
    \item \textbf{Training and Awareness:}  
    Introduce \textbf{continuous security awareness training} and \textbf{periodic phishing simulations} for employees.
    \item \textbf{Incident Response Plan:}  
    Define a procedure for \textbf{reporting suspicious emails}, including the use of automated phishing reporting tools.
    \item \textbf{Monitoring and Compliance:}  
    Communicate that \textbf{email activity may be monitored}, ensuring compliance with XYZ Bank’s security framework.
    \item \textbf{Continuous Evaluation and Policy Updates:}  
    Establish a mechanism for \textbf{regular policy reviews} to adapt to evolving cyber threats.
\end{itemize}

\subsubsection{Part 3: Reflection on AI vs. Human-Generated Policy Development (300 Words)}

Reflect on the process of using AI to draft an email policy versus developing your own research-based policy. Address the following:
\begin{enumerate}
    \item How did the AI-generated policy differ from your refined policy?  
    \item What were the limitations of the AI-generated policy in addressing \textbf{real-world risks and regulatory requirements}?  
    \item How did human expertise improve policy effectiveness and contextual relevance?  
    \item What role should AI play in \textbf{cybersecurity policy development}?  
\end{enumerate}

\subsection{Submission Requirements}
\begin{itemize}
    \item \textbf{Part 1:} AI-generated email policy (copy and paste output).  
    \item \textbf{Part 2:} Revised research-based email policy (500-800 words).  
    \item \textbf{Part 3:} Reflection on AI policy generation vs. human refinement (300 words).  
\end{itemize}

\end{appendices}


\bibliography{AI}% common bib file

\end{document}